\documentclass[letter,scriptaddress,twocolumn, showkeys]{revtex4}
\usepackage{ amssymb }
	\usepackage{amsmath}
	\usepackage{makeidx}
	\usepackage{amsfonts}
	\usepackage[ansinew]{inputenc}
	\usepackage[usenames,dvipsnames]{pstricks}
	\usepackage{courier}
	\usepackage{mathtools}
	\usepackage{subfigure}
	\usepackage{epsfig}
	\usepackage{pst-grad} 
	\usepackage{pst-plot} 
	\usepackage[colorlinks,hyperindex]{hyperref}
	\hypersetup
	{
		colorlinks,%
		citecolor=blue,%
		linkcolor=blue,%
		urlcolor=blue,%
	}

	\newtheorem{prop}{Proposition}
	\newtheorem{defn}{Definition}

\makeindex

\begin{document}

\title{On the geometry of regular icosahedral capsids \linebreak containing disymmetrons}

\author{Kai-Siang Ang $^{(a)}$
  and    Laura P. Schaposnik $^{(b)}$\\~}

    \affiliation{(a) The Harker School,    San Jose, CA 95128, USA\\
    (b)  University of Illinois, Chicago, IL 60607, USA}



\begin{abstract}
Icosahedral virus capsids are composed of symmetrons,  organized arrangements of capsomers. There are three types of symmetrons: disymmetrons, trisymmetrons, and pentasymmetrons, which have different shapes and are centered on the icosahedral 2-fold, 3-fold and 5-fold axes of symmetry, respectively. In 2010 [Sinkovits \& Baker] gave a classification of all possible ways of building an icosahedral structure solely from trisymmetrons and pentasymmetrons, which requires the triangulation number $T$ to be odd. In the present paper we incorporate disymmetrons to obtain a geometric classification of icosahedral viruses formed by regular penta-, tri-, and disymmetrons. For every class of solutions, we further provide formulas for symmetron sizes and parity restrictions on $h$, $k$, and $T$ numbers. We also present several methods in which invariants may be used to classify a given configuration.
\end{abstract}

\maketitle


\section{Introduction} 
\label{Introduction}

The two essential components of a virus are the genetic material and the protein {\it capsid} that surrounds the DNA or RNA. In 1956 Watson and Crick noticed that because the amount of genomic material in a virus could only encode for a few capsid proteins much smaller than the overall capsid, the genetic material must instead code for a few smaller proteins that are produced in large numbers and arranged symmetrically \cite{CW56}. The 60 symmetries of the icosahedron would then imply that there should be 60 proteins forming the capsid, but this is not what has been observed in most cases.
In 1962 Caspar and Klug used the idea of quasi-equivalence to explain how more than 60 proteins could come together \cite{CK62}. With 60 proteins, each face of the icosahedron has three proteins, one near each vertex. They proposed that the faces could be further triangulated, with each smaller triangle still having three proteins.

\begin{figure}[h]
\begin{center}
\includegraphics[width=0.45\textwidth]{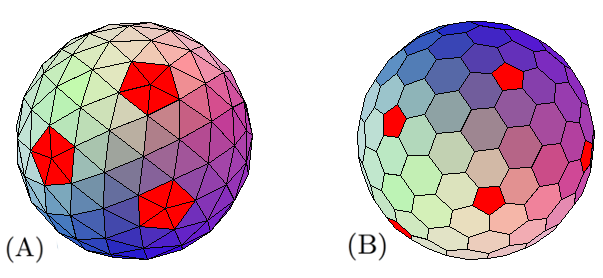}
\end{center}
\caption{Icosahedral capsid via the dual triangulated sphere, where 5-fold centers in  red and $(h,k)=(1,3)$. (A) Triangulated sphere; (B) dual space.}
\label{sphere}
\end{figure}

The triangulation can be seen on a triangulated sphere as in Figure \ref{sphere}.A. The original icosahedral vertices are at the centre of the red pentagons. The number of smaller triangles per original icosahedral face is known as the triangulation number $T$, given by the formula \cite{CK62} 
\begin{eqnarray}T=h^2+hk+k^2.\label{yo1}\end{eqnarray}  These $h,k$ parameters signify that if one starts from one 5-fold axis, walks $h$ steps in one direction along the triangulated sphere's edges, takes a $\frac{\pi}{3}$ turn left and walks $k$ steps, then the neighbouring 5-fold vertex will be reached.

The five or six proteins around a single vertex come together to form units called {\it capsomers}, which can be modelled as small spheres at the vertices produced by the triangulation. The total number of capsomers is related to the triangulation number by the formula  \cite{W69} \begin{eqnarray}N_{cap}=12 + 10(T-1).\label{yo2}\end{eqnarray} 

\begin{figure}[h]
\begin{center}
\includegraphics[width=0.4\textwidth]{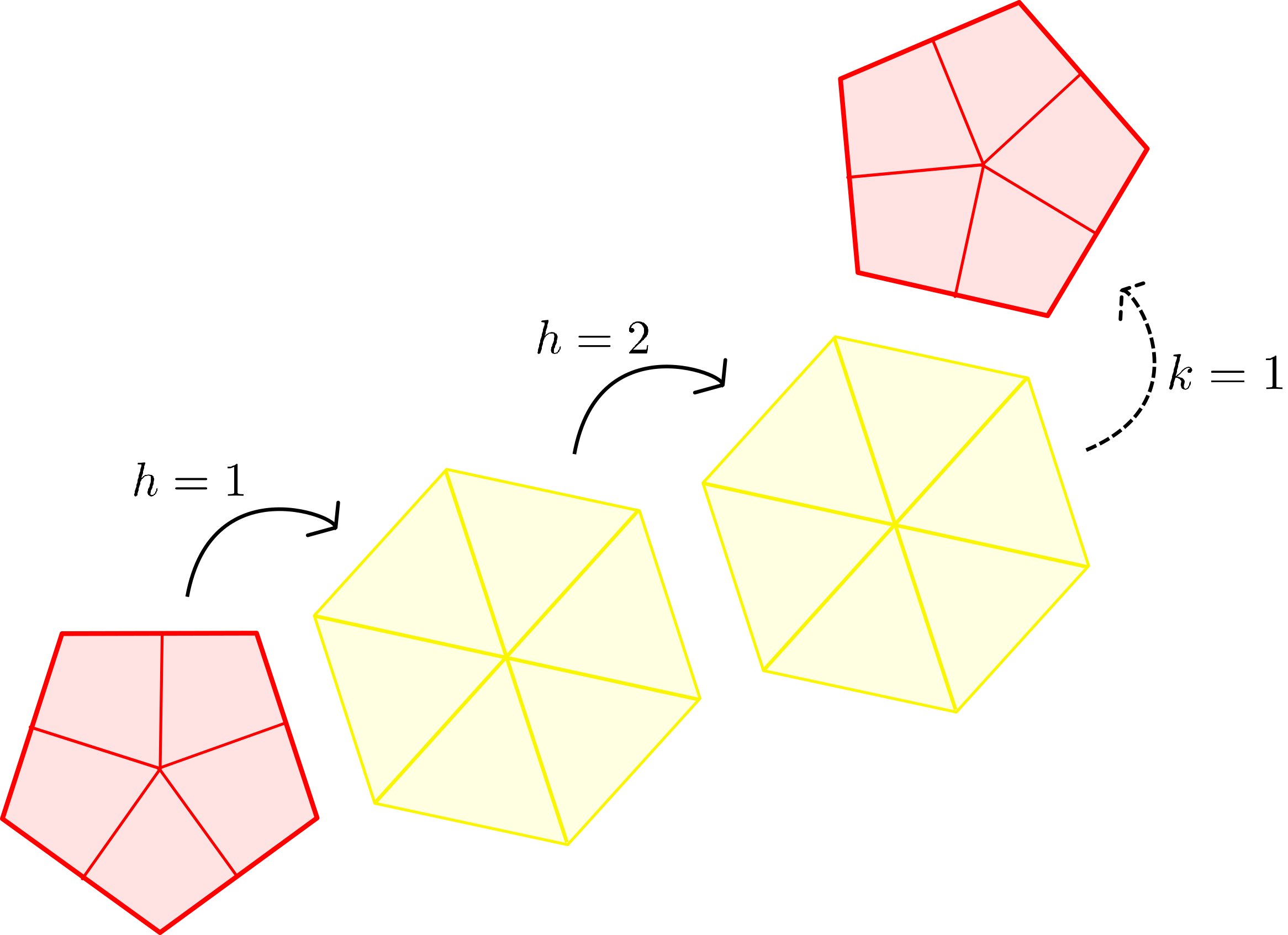}
 \end{center}
\caption{Triangulation number $T=7$ with $h=2$ and $k=1$.}
\label{steps}
\end{figure}

In Figure \ref{sphere}.B one can see the dual of the triangulated sphere, in which capsomers are represented by the pentagons or hexagons. Along the paper we shall use the triangulated sphere representation, as we can convert the 3D surface into a 2D coordinate system as follows: cut out and flatten a single icosahedral face that is triangulated. Then rotate this face about one of its vertices as in Figure \ref{steps}. With repeated rotations, one can tile the plane symmetrically, and the triangulation will provide the lines of a net of equilateral triangles.

\begin{figure}[h]
\begin{center}
\includegraphics[width=0.2\textwidth]{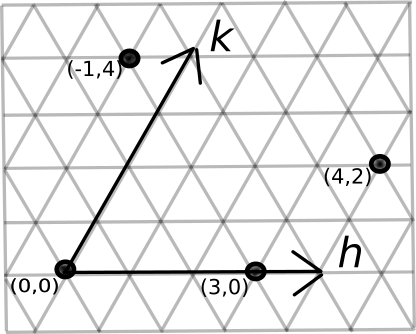}
 \end{center}
\caption{The triangular lattice with $h$- and $k$-axes. Examples are also given of some points and their coordinates.}
\label{grid}
\end{figure}

To reverse the process, one simply takes the triangle formed by three adjacent 6-fold centers of symmetry and map this triangle onto each face of the icosahedron.
  Since capsomers are located on vertices of the grid, they can be described through lattice points given by coordinates in the $h$ and $k$-axes. In polar coordinates, the $h$-axis is oriented along $\theta=0$ and the $k$-axis is oriented along $\theta=\frac{\pi}{3}$. An example is shown in Figure \ref{grid}. Note that pentasymmetrons become hexagons in 2-dimensions.

In 1969, Wrigley noticed in  \cite{W69} that virus capsids tend to dissociate into certain collections of capsomers called {\it symmetrons}; they had the shapes of regular pentagons ({\it pentasymmetrons}), equilateral triangles ({\it trisymmetrons}), and line segments ({\it disymmetrons}). Each of these symmetrons is centered on a corresponding axis of symmetry of the icosahedron (e.g., a pentasymmetron must be centered on an icosahedral vertex, through which a 5-fold axis of symmetry passes) \cite{W69}. Additionally, they must be arranged in a way that conforms exactly to the symmetry of the icosahedron. Due to their shapes,   symmetrons can be characterized simply by their edge lengths which we shall denote by $d$, $t$, and $p$ corresponding to di-, tri-, and pentasymmetrons respectively (these lengths are often denoted by $e_{DS}$, $e_{TS}$, and $e_{PS}$ in the literature). Since the icosahedral vertices are always  occupied by a pentasymmetron capsomer, one necessarily has $p>0$. Moreover, the symmetron shapes also lead to simple formulas for the number of capsomers per symmetron:
\begin{eqnarray}
N_{DS}&=&d, \label{yo}\\
N_{TS}&=&\frac{t(t+1)}{2}, \\
N_{PS}&=&1+\frac{5p(p-1)}{2}. \label{yo1}
\end{eqnarray}

The number of capsomers in each symmetron depends on its edge length, and 
the symmetries of the icosahedron determine how many of each symmetrons there are. The above equations \eqref{yo}-\eqref{yo1} can be combined with Wrigley's formula \eqref{yo2} relating the total number of symmetrons \linebreak$N_{cap}=N_{DS}+N_{TS}+N_{PS}$ to $T$ to produce the equation 
\begin{eqnarray}
T-1=3p(p-1)+t(t+1)+3d.
\end{eqnarray}

In 2010, Sinkovits and Baker gave a complete classification of all ways to arrange penta- and trisymmetrons to form the icosahedral capsid \cite{SB10}. However,  disymmetrons were not considered, and hence as noted by Wrigley, only accounted for viruses with even $T$ numbers \cite{W69}. We dedicate this paper to study the geometry of icosahedral viruses which include all three types of regular symmetrons, and give a classification of possible arrangements. Except for part of Section \ref{old}, where we recall Sinkovits and Baker's work \cite{SB10},  all that follows is original work.

\section{Towards a classification of icosahedral viruses}

In what follows we shall describe how icosahedral viruses can be classified by analysing the disymmetrons contained in the capsid, leading to 6 different classes of viruses. In the absence of disymmetron, i.e. when $d=0$, one recovers the results of \cite{SB10}, which we shall refer to as Class 1. In order to classify viruses for which $d>0$, we shall begin by defining what {\it bordering capsomers}.
\begin{defn}
A lattice point $(x,y)$ is {\bf adjacent} to another lattice point, if it is connected to it by exactly one edge in the lattice. 
\end{defn}
In the lattice representation a di-, tri-, and penta-symmetron is a collection of lattice points which form a line, a triangle or pentagon. From the above definition, the point $(x,y)$ is  adjacent to   $(x+1,y)$, $(x,y+1)$, $(x-1,y+1)$, $(x-1,y)$,  $(x,y-1)$, and $(x+1,y-1)$. 
\begin{figure}[h]
\begin{center}
\includegraphics[width=0.2\textwidth]{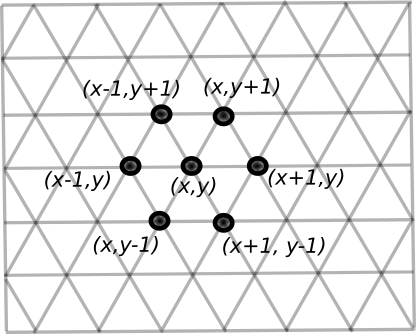}
 \end{center}
\caption{Lattice point $(x,y)$ and its 6 adjacent points.}
\end{figure}

\begin{defn}
In a symmetron $S$, a capsomer is an {\bf interior capsomer} if all adjacent lattice points are occupied by capsomers of $S$. Otherwise, the capsomer is an {\bf edge capsomer}.
\end{defn}

\begin{figure}[h]
\begin{center}
\includegraphics[width=0.2\textwidth]{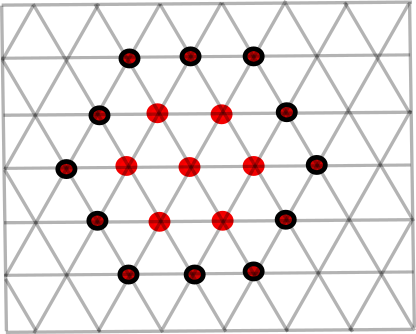}
 \end{center}
\caption{An example of a pentasymmetron with edge capsomers (black) distinguished from interior capsomers (red).}
\end{figure}

 \pagebreak

\begin{defn}
Set the grid so that a disymmetron $D$   lies along the line $k=0$. Any $k=\pm1$ capsomer adjacent to a   capsomer of $D$ is a {\bf bordering capsomer} with respect to $D$.
\end{defn}

\begin{figure}[h]
\begin{center}
\includegraphics[width=0.2\textwidth]{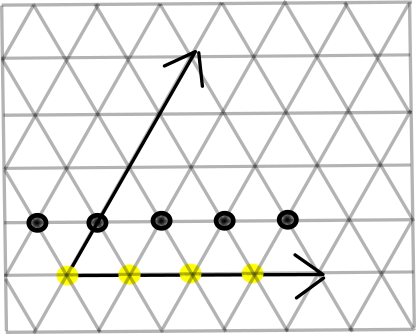}
 \end{center}
\caption{An example of a disymmetron of length 4 (yellow) and its 5 (black) $k=1$ bordering capsomers.}
\end{figure}

\begin{defn}
A symmetron is a {\bf bordering symmetron} that {\bf borders} the disymmetron $D$ if either:
\begin{itemize}
\item it has at least two capsomers with the same $k$ coordinate which are bordering capsomers with respect to $D$.
\item or, all of its capsomers are bordering capsomers with respect to $D$.
\end{itemize}
The $k$ coordinate of the bordering capsomers determines which {\bf side} of $D$ the symmetron is bordering.
\end{defn}

\begin{figure}[h]
\begin{center}
\includegraphics[width=0.2\textwidth]{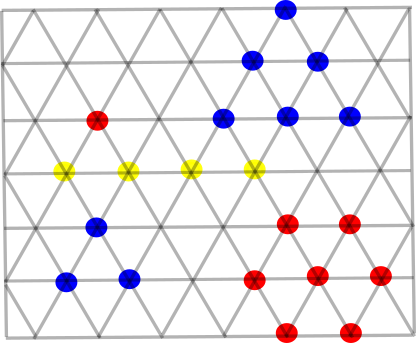}
 \end{center}
\caption{A disymmetron (yellow) and above it, its bordering pentasymmetron (red) and trisymmetron (blue). The   penta- and trisymmetron below the disymmetron are not bordering, since neither of the items in the definition apply.}
\end{figure}

As can be seen from the above definitions, bordering always occurs with respect to a disymmetron, so for simplicity we shall write {\it pentasymmetrons border} when referring to {\it pentasymmetrons bordering disymmetrons}.  Because of the convex shape of penta- and trisymmetrons, a bordering penta- or trisymmetron will have a single string of edge capsomers bordering exactly one side of the disymmetron. The symmetron shapes also dictate that if a symmetron $S$ has bordering capsomers with $k=1$, then all of the capsomers of $S$ have $k\ge 1$. Moreover,  2-fold symmetry about the disymmetron dictates that another symmetron $S'$ must be bordering the disymmetron on the opposite side. From the above, on each side of the disymmetron there can be at most one bordering di-, at most one bordering tri-, and at most one bordering pentasymmetron.

By a \textit{symmetry of overlapping argument} (through which one applies icosahedral symmetry to find an overlap of symmetrons, giving a proof by contradiction), a disymmetron centered on an edge $e$ can only be adjacent to the 4 disymmetrons centered on edges belonging to icosahedral faces with edge $e$, the 2 trisymmetrons centered on icosahedral faces with edge $e$, and the 2 pentasymmetrons centered on the endpoints of edge $e$. Note that two disymmetrons bordering each other must be parallel, but if they are centered on two edges of a single icosahedral face $F$, then this is not possible if $d>1$ (3-fold symmetry about the center of $F$ dictates that they must be at $\frac{\pi}{3}$ to each other). Therefore disymmetrons can not border each other unless $d=1$.
In Appendix \ref{Ap3}, we prove that bordering symmetrons must exist if disymmetrons exist, i.e., when  $d>0$.

 In what follows we shall use geometric methods to give a classification of icosahedral viruses based upon what types of symmetrons border the disymmetrons leading to six distinct classes.  When only pentasymmetrons border, one has Classes 2 and 5; when only trisymmetrons border, one has Classes 3 and 6; when both tri- and pentasymmetrons border, one has Class 4. As stated before, when $d=0$ gives Class 1. We shall now show how these Classes arise and prove that they are the only possible configurations. Unless stated otherwise, we shall set the coordinate grid so that, if the disymmetron exists, it occupies the lattice points from $(1,0)$ to $(d,0)$. We shall also be applying 2-, 3-, or 6-fold symmetries, implicitly referring to the formulas given in Appendix \ref{Ap1}.

\subsection{Class 1: No disymmetrons, $d=0$}\label{old}

Sinkovits and Baker showed that there are 3 configurations for $d=0$, and we refer the reader to  \cite{SB10} for the proof that these are all that exist. These authors considered three different configurations because several different formulas arise (we will derive them in Section \ref{formulas}) and the distinction was necessary to conform to their restriction that $h\le k$.

Along this paper, we drop the restriction $h\le k$, and thus treat all 3 configurations of  \cite{SB10} as the same. Note that handedness can be accounted for with a reflection of the plane, or a switch between $h$ and $k$ coordinates. What these configurations have in common is that, after a reflection of the plane if needed, one can set a coordinate grid so that:
\begin{itemize}
\item There is a pentasymmetron $P_1$ with center at $(0,0)$.
\item $P_1$ has an edge from $(1-p,p-1)$ to $(0,p-1)$.
\item There is a trisymmetron edge from $(1-p,p)$ to $(t-p,p)$.
\item There is a pentasymmetron $P_2$ with an edge from $(t-p+1,p)$ to $(t,p)$.
\item $P_2$ has its center at $(t-p+1,2p-1)$.
\end{itemize}
This information, which we call the  Grid Description {(GD)}, combined with the icosahedral symmetries, is sufficient to fill the plane. Examples can be seen in the following Figure \ref{fig1}.

\begin{figure}[h]
\begin{center}
\includegraphics[width=0.45\textwidth]{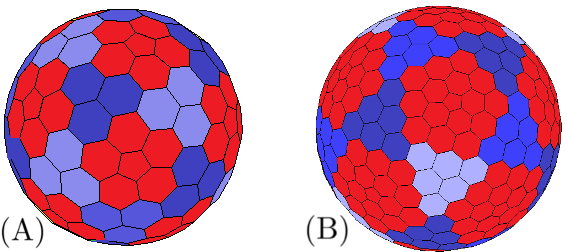}
 \end{center}
\caption{Examples of Class 1 viruses: (A) no disymmetrons, corresponding to $d=0$, $t=2$, and $p=2$ and $(h,k)=(1,3)$; (B) no disymmetrons, corresponding to $d=0$, $t=3$, and $p=3$ and $(h,k)=(5,1)$. Different shades of blue are used to distinguish different trisymmetrons.}
\label{fig1}
\end{figure}
%

In what follows we shall describe the five classes for which $d>0$ by giving its GD and showing why this description provides sufficient information to the determine the class of a virus. 
 
\subsection{Class 2: Only pentasymmetrons border} \label{cl2}

\begin{figure}[h]
\begin{center}
\includegraphics[width=0.45\textwidth]{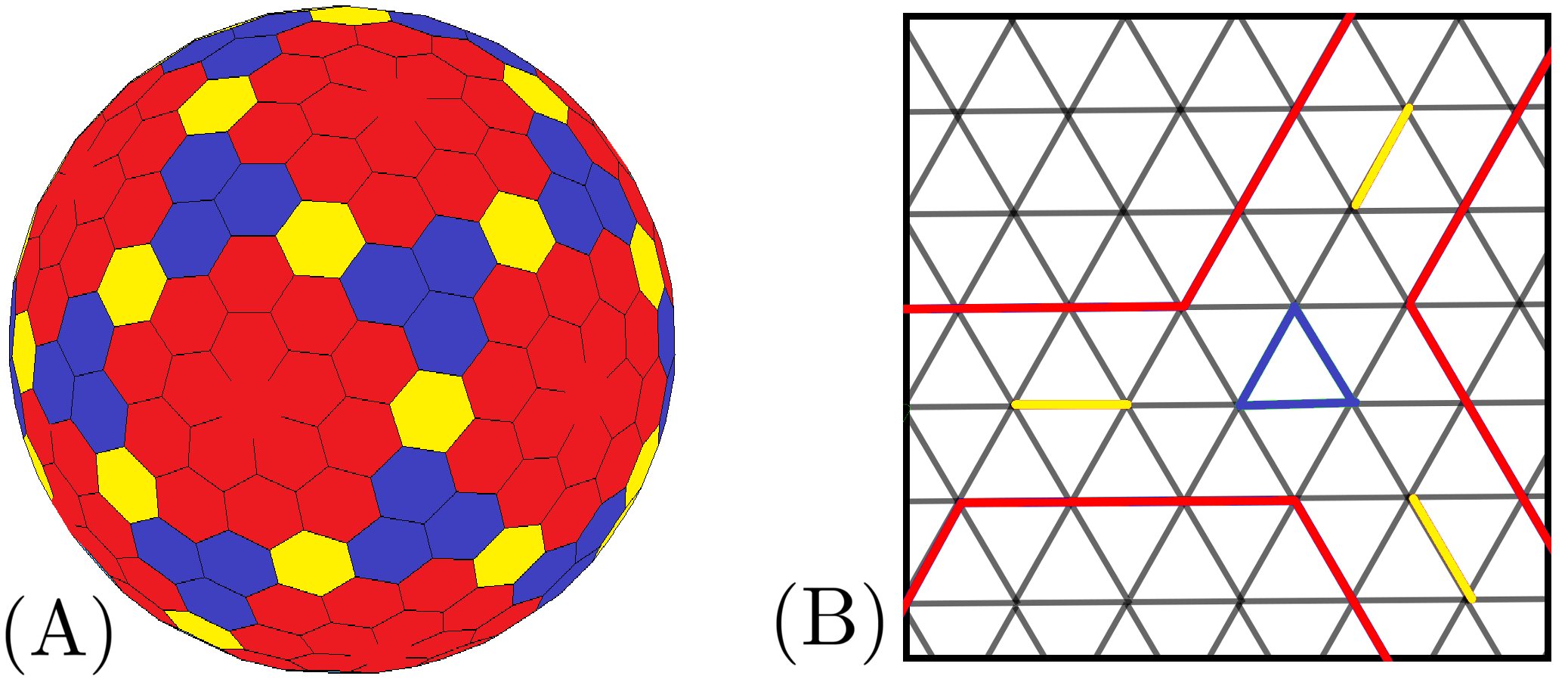}
 \end{center}
\caption{Examples of viruses in Class 2 where only pentasymmetrons border: (A) corresponding to $d=1$, $t=2$, and $p=3$ and $(h,k)=(2,4)$; (B) corresponding to $d=2$, $t=2$, and $p=4$ and $(h,k)=(3,5)$.}
\label{fig2}
\end{figure}

This Class has the following Grid Description:
\begin{itemize}
\item There is a pentasymmetron $P_1$ with center at $(0,0)$.
\item $P_1$ has an edge from $(1-p,p-1)$ to $(0,p-1)$.
\item There is a disymmetron from $(1,p-1)$ to $(d,p-d)$.
\item There is a disymmetron from $(1-p,p)$ to $(d-p,p)$.
\item There is a trisymmetron from $(d-p+1,p)$ to $(0,p)$.
\item There is a pentasymmetron $P_2$ with an edge from $(d+1,p-d)$ to $(d+p,p-d)$.
\item $P_2$ has its center at $$(d+1,2p-d-1)=(p-t+1,p+t-1).$$
\end{itemize}
Examples can be seen in Figure \ref{fig2}. In order to see that the above GD determines the class,  consider a pentasymmetron that has an edge with $k=c$ for some constant $c$, and let all of its capsomers have $k\ge c$. Considering the 6-fold symmetry around the pentasymmetron, one can see that there will be overlapping capsomers if all of the $k=c-1$ lattice points adjacent to the $k=c$ edge capsomers are occupied by a single disymmetron. Now recall that we have set the grid so that there exists a disymmetron from $(1,0)$ to $(d,0)$. Without loss of generality, we may consider the bordering pentasymmetron that has $k=1$ capsomers from $(m,1)$ to $(n,1)$, with $m\le0\le n\le d-1$. Then the pentasymmetron center is at $(m,1+n-m)$.

Notice that $(1,0)$ gets mapped to $(n+1,1-m)$ under a $\frac{\pi}{3}$ rotation about the pentasymmetron center. The points from $(n+1,1)$ to $(n+1,-m)$ can not be pentasymmetron capsomers by Appendix \ref{pentatriadj} and can not be disymmetrons by 6-fold symmetry, so they must be trisymmetron capsomers. Each of the points are of the form $(n+1,p)$, with $1\le p\le-m$. By a $-\frac{\pi}{3}$ rotation about the pentasymmetron center, they are mapped to $(m+p,0)$.

If there are 0 such $(n+1,p)$ points, then $m=0$, and we can apply 2-fold and 6-fold symmetries to partially fill the grid. Note that if $n<d-2 $, then  $(n+2,1)$ can not be a penta- or disymmetron capsomer, so it must be a trisymmetron capsomer. But this would make both penta- and trisymmetrons border the disymmetron, which lies outside of this case. The case of $n=d-1$ fixes $t=0$, and there is enough information to see that this arrangement fits the GD and fills the plane. The case of $n=d-2$ also fixes $t=0$, and there is enough information to fill the plane. However, this arrangement does not conform to the GD and we shall see in  Section \ref{cl4}
 that it gives a degenerate form of Class 4.

We shall now assume that there is at least 1 point of the form $(n+1,p)$. These points belong to the same trisymmetron. If there are at least 2 of such points, the trisymmetron orientation becomes fixed. If there is only one  such point, the same orientation can occur, but an alternate orientation becomes possible and is covered in Section \ref{cl5}, not here. By 6-fold symmetry, there is another trisymmetron with edge along $(m+p,0)$. By 2-fold symmetry about the $k=0$ disymmetron, the trisymmetron with edge $(n+1,p)$ must also have an edge with $k=0$, fixing the size so that $d+t=p$. This is enough information to cover the plane and verify that this conforms to the GD.

\subsection{Class 3: Only trisymmetrons border} \label{cl3}

This Class has the following Grid Description:
\begin{itemize}
\item There is a pentasymmetron $P_1$ with center at $(0,0)$.
\item $P_1$ has an edge from $(1-p,p-1)$ to $(0,p-1)$.
\item There is a disymmetron from $(1,p-1)$ to $(d,p-1)$.
\item There is a trisymmetron from $(2-p,p)$ to $(d-1,p)$.
\item There is a pentasymmetron $P_2$ with an edge from $(d+1,p-1)$ to $(d+p,p-1)$.
\item $P_2$ has its center at $(d+1,2p-2)=(t-p+3,2p-2)$.
\end{itemize}

\begin{figure}[h]
\begin{center}
\includegraphics[width=0.45\textwidth]{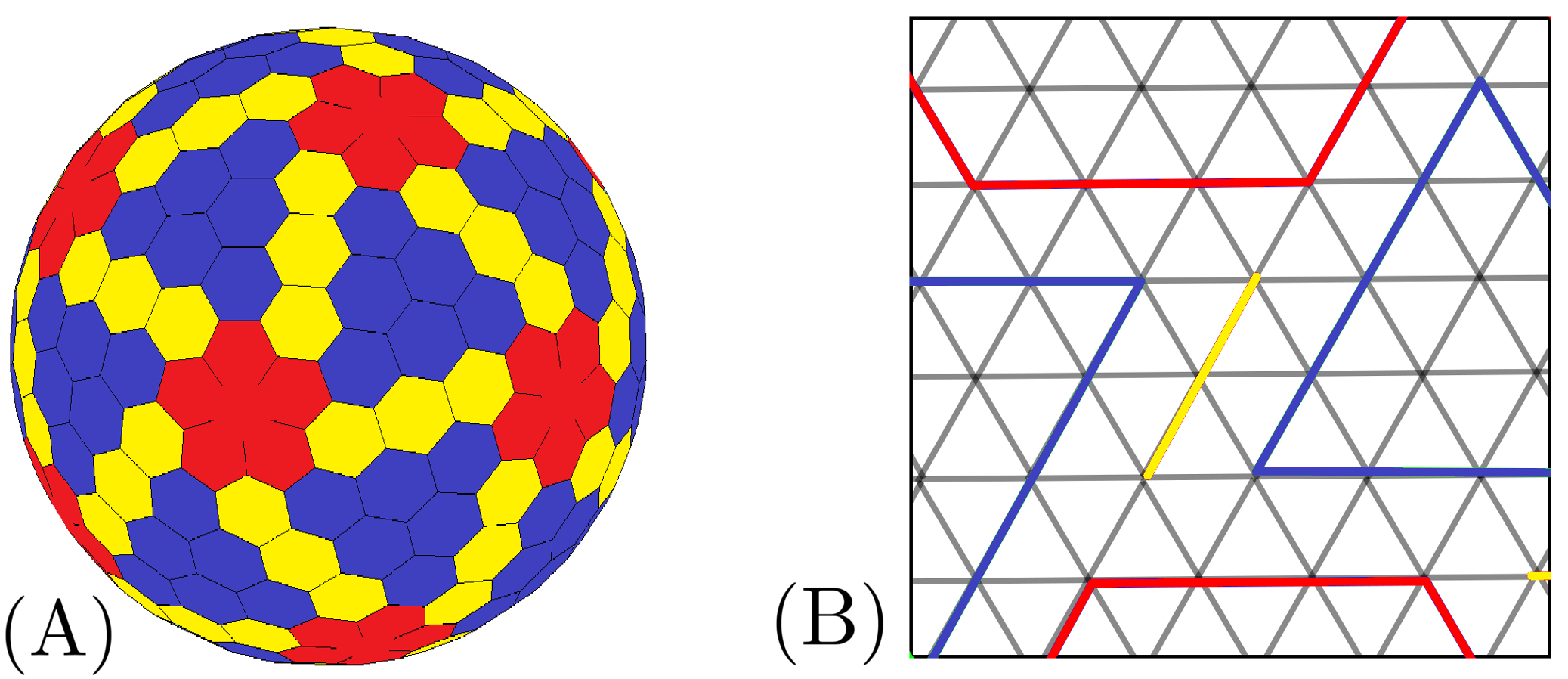}
 \end{center}
\caption{Examples of viruses in Class 3 where only trisymmetrons border: (A) corresponding to $d=3$, $t=3$,  $p=2$, and $(h,k)=(2,4)$; (B)  corresponding to $d=3$, $t=5$,  $p=4$, and $(h,k)=(4,6)$.}
\label{fig3}
\end{figure}

Examples can be seen in Figure \ref{fig3}.

We now show how this configuration arises. As in the previous section, without loss of generality, we consider the trisymmetron having $k=1$ capsomers from $(m,1)$ to $(n,1)$, with 
\begin{eqnarray}m\le0\le n.\label{mn}\end{eqnarray} 
The third vertex of the trisymmetron is at $(m,1+n-m)$, which makes the center $(\frac{2m+n}{3},\frac{3+n-m}{3})$.

By Appendix \ref{Ap1}, under $120^\circ$ rotation, $(1,0)$ goes to $(m+n,2-m)$. Since these points can not be tri- or disymmetron capsomers by Appendix \ref{Ap2} and 6-fold symmetry, the points from $(n+1,1)$ to $(m+n+1,1-m)$, which we shall call the $e$ points, must be pentasymmetron capsomers of the same pentasymmetron by  Appendix \ref{Ap2}.
 
Since $p>0$, there will always be at least one $e$ point. Note that if there are at least two $e$ points, then they form an edge that defines the orientation of the pentasymmetron. This orientation may occur if there is only one $e$ point, but there is also an alternate orientation as we  discuss in Section \ref{cl6}. Rotating the edge between $(n+1,1)$ and $(m+n+1,1-m)$ by $120^\circ$ clockwise, we see that a different pentasymmetron has a $k=0$ edge. By 2-fold symmetry around the $k=0$ disymmetron, the pentasymmetron containing $(n+1,1)$ must also have a $k=0$ edge, fixing the pentasymmetron edge size: $d+p=t+2$. This is enough information to fill the plane  and thus determines the class.

\subsection{Class 4: Both tri- and pentasymmetrons border} \label{cl4}

This Class has the following Grid Description:
\begin{itemize}
\item There is a pentasymmetron $P_1$ with center at $(0,0)$.
\item $P_1$ has an edge from $(1-p,p-1)$ to $(0,p-1)$.
\item There is a disymmetron from $(1-p,p)$ to $(d-p,p)$.
\item There is a trisymmetron edge from $(1-p,p+1)$ to $(t-p,p+1)$.
\item There is a pentasymmetron $P_2$ with an edge from $(d-2p+1,p+1)$ to $(d-p,p+1)$.
\item $P_2$ has its center at $(d-2p+1,2p)=(t-p+2,2p)$.
\end{itemize}

\begin{figure}[h]
\begin{center}
\includegraphics[width=0.45\textwidth]{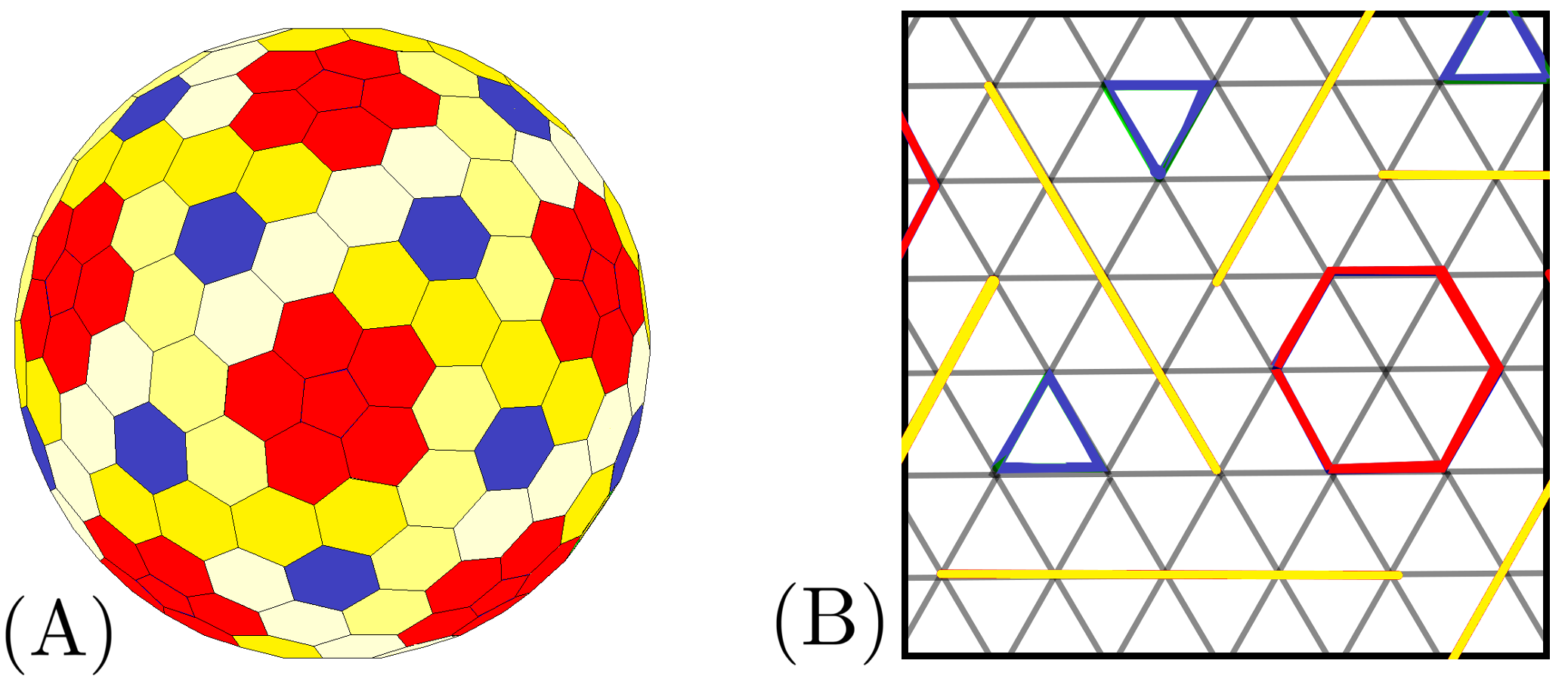}
 \end{center}
\caption{Examples of viruses in Class 4 where both tri- and pentasymmetrons border: (A) corresponding to $d=4$, $t=1$,  $p=2$, and $(h,k)=(1,4)$, where different shades of yellow are used to distinguish different disymmetrons; (B) corresponding to $d=5$, $t=2$,  $p=2$, and $(h,k)=(2,4)$.}
\label{fig4}
\end{figure}

Examples can be seen in Figure \ref{fig4}.\\

%

In this case, since both the penta- and trisymmetrons border the disymmetron, they both have no $k=0$ capsomers, so $(0,0)$ is a disymmetron capsomer. Suppose that a bordering trisymmetron is adjacent to $(0,0)$. It can not also be adjacent to $(d+1,0)$, since otherwise the pentasymmetron would not be able to border. Applying 3-fold symmetry, we see that there will be a disymmetron capsomer with $h<0$ and $k=0$, and this capsomer will be part of a disymmetron parallel to the $k$-axis. Thus the disymmetron containing $(0,0)$ must be parallel to $h+k=0$. However, it will not be bordering either of the trisymmetrons bordering the $k=0$ disymmetron. Hence, the bordering trisymmetron cannot be adjacent to any non-disymmetron $k=0$ capsomers, and each edge of the trisymmetron is completely bordering a disymmetron.

Consider now all of the $k=1$ bordering capsomers. Since 1 penta- and 1 trisymmetron border a given  side of the disymmetron, there will be exactly 1 string of pentasymmetron capsomers and 1 string of trisymmetron capsomers. A total of 4 disymmetrons may be adjacent to but not bordering the $k=0$ disymmetron, and 2 of these disymmetrons may each have at most one adjacent $k=1$ capsomer. Because each edge of the trisymmetron completely borders a disymmetron, on either side of the string of the trisymmetron $k=1$ bordering capsomers, there will be a disymmetron capsomer. The only other component of the $k=1$ bordering capsomers is the string of pentasymmetron bordering capsomers, which is immediately before or after the string of di-tri-disymmetron capsomers.

In view of the analysis above and   without loss of generality, we may assume that a trisymmetron edge goes from $(d-1,1)$ to $(d-t,1)$. This means that the third trisymmetron vertex is at $(d-t,t)$, so the trisymmetron center is located at $(\frac{3d-2t-1}{3},\frac{t+2}{3})$. 3-fold symmetry means a disymmetron must exist from $(d-t-1,1)$ to $(d-t-1,d)$. By 2-fold symmetry about the $k=0$ disymmetron, there must be a trisymmetron with edge from $(2,-1)$ to $(1+t,-1)$ and third vertex at $(1+t,-t)$. By 3-fold symmetry about this new trisymmetron, there is a disymmetron from $(1+t,-1-t)$ to $(2+t-d,-2-t+d)$. 

Consider the $k=1$ capsomers adjacent to the $k=0$ disymmetron. For a pentasymmetron to border the disymmetron, $d\ge t+3$. Therefore the disymmetron from $(1+t,-1-t)$ to $(2+t-d,-2-t+d)$ passes through $(-1,1)$. Consider the lattice points from $(0,1)$ to $(d-t-2,1)$. These must belong to a single pentasymmetron, since no two symmetrons of the same type may be bordering the same side of a disymmetron. This fixes the pentasymmetron orientation and gives $d=t+p+1$, thus completing the plane.
\subsection{Class 5: An exceptional case when only pentasymmetrons border}\label{cl5}

This Class has the following GD:
\begin{itemize}
\item There is a pentasymmetron $P_1$ with center at $(0,0)$.
\item $P_1$ has an edge from $(1-p,p-1)$ to $(0,p-1)$.
\item There is a disymmetron from $(0,p)$ to $(p-2,2)$.
\item There is a trisymmetron edge from $(-1,p)$ to $(-1,p+1)$.
\item There is a pentasymmetron $P_2$ with an edge from $(p-2,3)$ to $(2p-3,3)$.
\item $P_2$ has its center at $(p-2,p+2)$.
\end{itemize}

\begin{figure}[h]
\begin{center}
\includegraphics[width=0.45\textwidth]{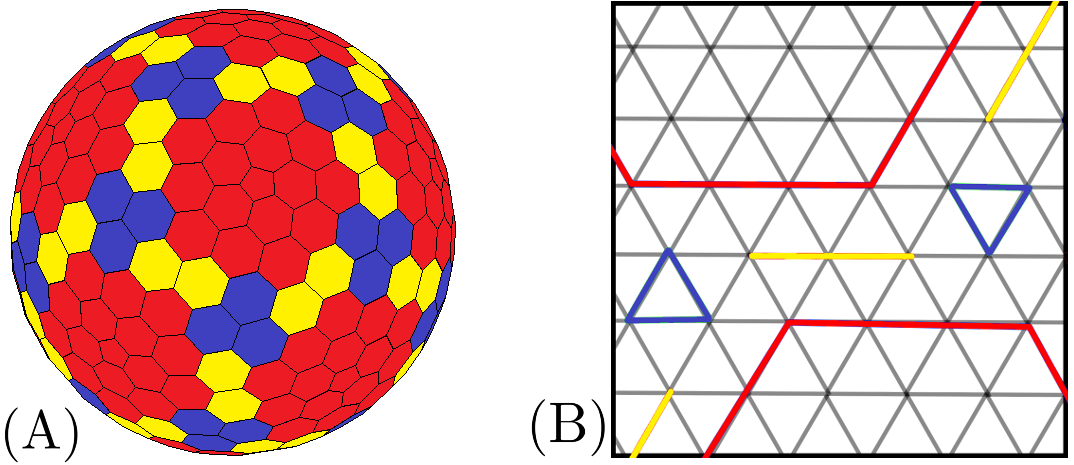}
 \end{center}
\caption{Examples of viruses in Class 5 giving exceptional case of when only pentasymmetrons border: (A) corresponding to $d=2$, $t=2$,  $p=3$, and $(h,k)=(5,1)$; (B) corresponding to $d=3$, $t=2$,  $p=4$, and $(h,k)=(2,6)$.}
\label{fig5}
\end{figure}

Examples can be seen in Figure \ref{fig5}.\\

%

We now show how this configuration arises. In this case only pentasymmetrons border and there is only 1 $(n+1,p)$ point. In addition to the orientation covered in Section \ref{cl2}, there is another orientation in which the trisymmetron has an edge along $h=1$. Because there is another trisymmetron with a $k=0$ vertex capsomer, 2-fold symmetry about the $k=0$ disymmetron forces the trisymmetron containing $(n+1,1)$ to have a vertex capsomer with $k=0$. This fixes $t=2$, and the point $(n+1,0)$ can now only be a disymmetron, fixing the disymmetron size so that we have $p=d+1$, and therefore the plane can be canonically filled. 
\subsection{Class 6: An exceptional case when only trisymmetrons border} \label{cl6}

This Class has the following GD:
\begin{itemize}
\item There is a pentasymmetron $P_1$ with center at $(0,0)$.
\item $P_1$ has an edge from $(-1,1)$ to $(0,1)$.
\item There is a disymmetron from $(0,2)$ to $(0,d+1)$.
\item There is a trisymmetron edge from $(-1,2)$ to $(-1,d+2)$.
\item There is a pentasymmetron $P_2$ with an edge from $(0,d+2)$ to $(1,d+2)$.
\item $P_2$ has its center at $(0,d+3)=(0,t+2)$.
\end{itemize}

\begin{figure}[h]
\begin{center}
\includegraphics[width=0.45\textwidth]{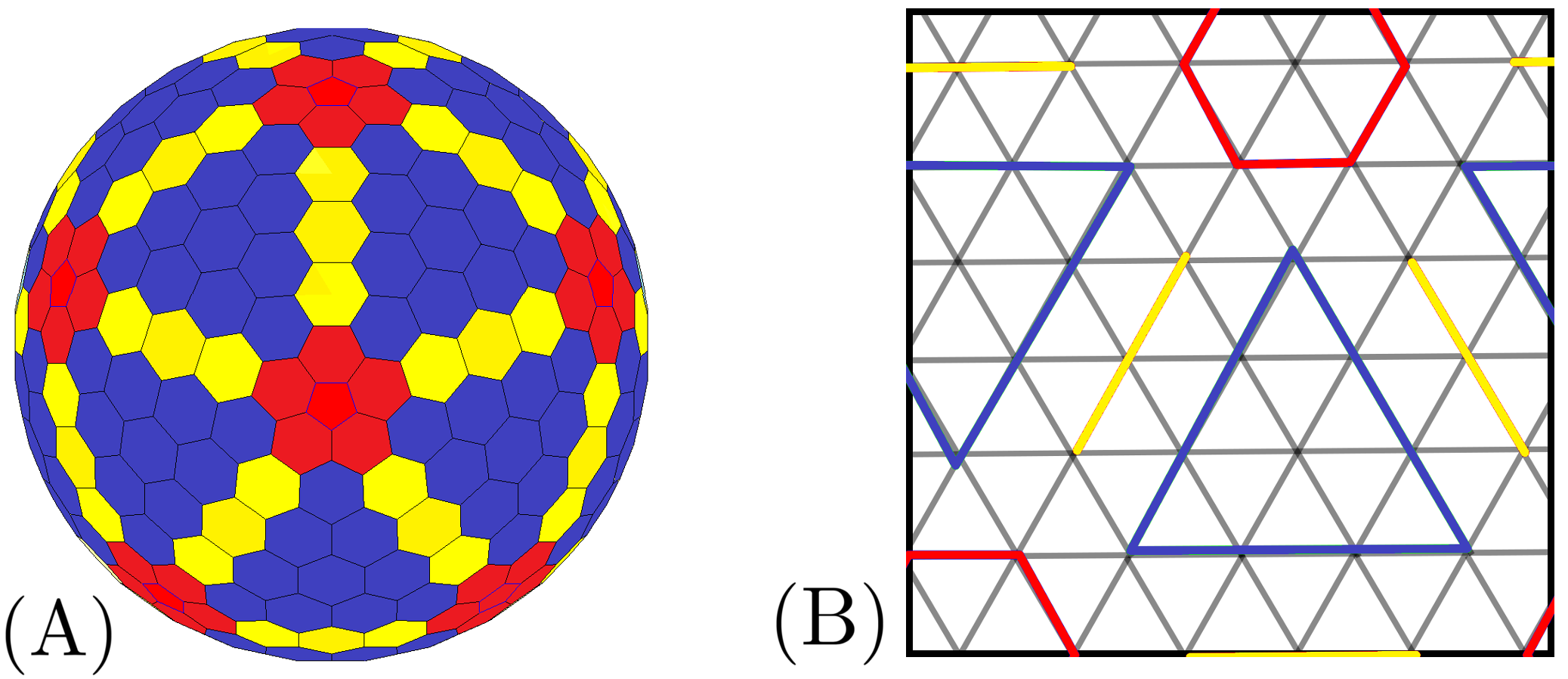}
 \end{center}
\caption{Examples of viruses in Class 6 giving exceptional case of when only trisymmetrons border: (A) corresponding to $d=3$, $t=4$, $p=2$, and $(h,k)=(0,6)$; (B) corresponding to $d=3$, $t=4$, $p=2$, and $(h,k)=(0,6)$.}
\label{fig6}
\end{figure}

Examples can be seen in Figure \ref{fig6}.\\

%

 This is the case in which only trisymmetrons border and there is only 1 pentasymmetron capsomer among the $e$ points mentioned in Section \ref{cl3}, implying $m=0$ in \eqref{mn}. In addition to the orientation covered in that section, there is another orientation in which the pentasymmetron has an edge with $k=1$. The disymmetron containing $(n,2)$ has a pentasymmetron vertex capsomer at one end, at $(n+1,1)$. Therefore, the $k=0$ disymmetron must have pentasymmetron vertex capsomers at its ends also. Note that these ends also have $k=0$. If the pentasymmetron has a $k=1$ edge, then the only way for it to have a $k=0$ vertex capsomer is to have one at $(n+1,0)$. This gives $p=2$ and $d+1=t$, and   the plane can be filled.

\section{Formulas and Invariants} \label{formulas}

As in ref.~\cite{SB10}, we now derive formulas for $d$, $t$, and $p$ based on $h$ and $k$, and we also provide parity restrictions on $h$ and $k$.
In our Grid Descriptions, we examined the position of the centre of      adjacent pentasymmetrons  with respect to a pentasymmetron centered at $(0,0)$, which leads  to the following table:

\begin{table}[h]
  \begin{tabular}{ |c| c | c |}\hline
     Class & Coordinates & Extra Equations\\ \hline
     1 & $(t-p+1,2p-1)$ & $d=0$\\\hline
     2 & $(p-t+1,p+t-1)$ & $p=d+t$\\\hline
     3 & $(t-p+3,2p-2)$ & $t+2=p+d$\\\hline
     4 & $(t-p+2,2p)$ & $t+p+1=d$\\\hline
     5 & $(p-2,p+2)$ & $t=2$, $d+1=p$\\\hline
     6 & $(0,t+2)$ & $p=2$, $d+1=t$\\\hline
  \end{tabular}
   \caption{Coordinates of adjacent pentasymmetron centers with respect to a pentasymmetron at the origin.}
   \label{coords}
\end{table}

Note that $p>0$ and (with the exception of Class 1), $d>0$. For all of the coordinate values, the $k$ coordinate is non-negative. Also, for Classes 2, 3, 5, and 6, we can use the extra equations to see that also $h \ge0$. The reason we are concerned with the  signs of the coordinates $h,k$ is  because   the $(h,k)$ characterisation of the icosahedral triangulation  requires that $h,k\ge0$. 

 If the coordinate of the adjacent pentasymmetron centre has $0^\circ\le\theta\le\frac{\pi}{3}$, then the $h$ steps taken in the $(h,k)$ characterisation will be in the direction of $\theta=0^\circ$ (Case A). If the coordinate has $\frac{\pi}{3}\le\theta\le120^\circ$, then the $h$ steps are in the direction of $\theta=\frac{\pi}{3}$ (Case B). If the coordinate has $120^\circ\le\theta\le180^\circ$, then the $h$ steps are in the direction of $\theta=120^\circ$ (Case C). Since the $k$ coordinates of our adjacent pentasymmetron centers are always non-negative, one does not need to consider $\theta>180^\circ$. Finally, when $h \ge0$,  one has Case A. If we do not have these restrictions on $h$, we may also have Cases B or C.

We can find formulas of the $(h,k)$ characterisation for the Case A solely based on the coordinates of the adjacent pentasymmetron center. In Case B, the coordinates of the adjacent pentasymmetron center must be rotated clockwise by $\frac{\pi}{3}$, i.e. $(x,y)\to(x+y,-x)$. In Case C, the coordinates must be rotated clockwise by $120^\circ$, given by $(x,y)\to(y,-x-y)$. Using these transformations and the coordinates listed in Table \ref{coords}, we can solve for $d$, $t$, and $p$ based on $(h,k)$. Additionally, we need to reverse $h$ and $k$ in each equation to find solutions in a flipped orientation. One should also note that $d$, $t$, and $p$ must be integral, giving parity restrictions on $h$ and $k$, which in turn give parity restrictions on $T$ (since $T=h^2+hk+k^2$), leading to  Table \ref{dtphkT}.

 In what follows we shall describe different methods through which one can  distinguish the class of an icosahedral virus. These are:

 \begin{itemize}
\item To check whether a configuration conforms to each Grid Description.
\item To do a visual classification: look for the existence of disymmetrons and, for each pair of types of symmetrons, look at whether these symmetrons' closest edges are parallel or not. This is slightly error-prone.
\item To substitute the specific values for $h$ and $k$ into Table \ref{dtphkT} and see which Class yields the correct values of $d$, $t$, and $p$.
\item To use the numerical test we describe below.
\end{itemize}

\begin{table}[h]
\begin{tabular}{ |c|c|c|c|c|c| } 
 \hline
 Class & $d$     & $t$  & $p$ & $h,k$ Parity & $T$ Parity\\
 \hline
 1 & $0$  & $\frac{h-1}{2}+k$& $\frac{h+1}{2}$ & $h\equiv 1$ & $T\equiv 1$\\
 \hline
 1 & $0$  & $\frac{k-1}{2}+h$& $\frac{k+1}{2}$ & $k\equiv 1$ & $T\equiv 1$\\
 \hline
 1 & $0$  & $\frac{k-h-1}{2}$& $\frac{h+k+1}{2}$ & $h\not\equiv k$ & $T\equiv 1$\\
 \hline
 1 & $0$  & $\frac{h-k-1}{2}$& $\frac{h+k+1}{2}$ & $h\not\equiv k$ & $T\equiv 1$\\
 \hline
 2 & $h-1$  & $\frac{k-h}{2}+1$ & $\frac{h+k}{2}$ & $h\equiv k$ & $T\equiv h$\\
 \hline
 2 & $k-1$  & $\frac{h-k}{2}+1$ & $\frac{h+k}{2}$ & $h\equiv k$ & $T\equiv h$\\
 \hline
 3 & $h-1$  & $h+\frac{k}{2}-2$& $\frac{k}{2}+1$ & $k\equiv0$ &$ T\equiv h$\\
  \hline
 3 & $k-1$ & $\frac{h}{2}+k-2$& $\frac{h}{2}+1$  & $h\equiv0$ & $T\equiv k$\\
 \hline
 4 & $h+k-1$ & $h+\frac{k}{2}-2$& $\frac{k}{2}$  & $k\equiv0$ & $T\equiv h$\\
 \hline
 4 & $h+k-1$  & $\frac{h}{2}+k-2$& $\frac{h}{2}$ & $h\equiv0$ & $T\equiv k$\\
 \hline
 4 & $k-1$  & $\frac{k-h}{2}-2$& $\frac{h+k}{2}$ & $h\equiv k$ & $T\equiv h$\\
 \hline
 4 & $h-1$  & $\frac{h-k}{2}-2$& $\frac{h+k}{2}$ & $h\equiv k$ & $T\equiv h$\\
 \hline
 5 & $h+1$ & $2$& $h+2$  & $k=h+4$ & $T\equiv h$\\
 \hline
 5 & $k+1$ & $2$& $k+2$  & $h=k+4$ & $T\equiv h$\\
 \hline
 6 & $k-3$ & $k-2$& $2$  & $h=0$ & $T\equiv k$\\
 \hline
 6 & $h-3$ & $h-2$& $2$  & $k=0$ & $T\equiv h$\\
 \hline
\end{tabular}
\caption{Formulas for $d$, $t$, and $p$ in terms of $h$ and $k$, as well as parity restrictions on $h$, $k$, and $T$.}
   \label{dtphkT}
 \end{table}

As explained in Appendix \ref{Ap4}, the icosahedral edge length squared is equal to $T$. Also from Appendix \ref{Ap4}, we can use the coordinates from Table \ref{coords} to find this edge length squared in terms of $p$ and $t$. This motivates the following pseudo-code test, which takes inputs $d$, $t$, $p$, and $T$, and outputs the Class number:\\

\begin{small}
\texttt{if ($d$==0) return 1;}

\texttt{else \{}

\texttt{\quad if (3$p^2$+$t^2$-2$t$+1==$T$) return 2;}

\texttt{\quad if (3$p^2$-6$p$+$t^2$+4$t$+7==$T$ \&\& $t$!=$p$-1) return 3;}

\texttt{\quad if (3$p^2$+$t^2$+4$t$+4==$T$) return 4;}

\texttt{\quad if (3$p^2$+4==$T$ \&\& $t$==2 \&\& $p$!=2) return 5;}

\texttt{\quad if ($t^2$+4$t$+4==$T$ \&\& $p$==2) return 6;}

\texttt{\}\\}
\end{small}

There are a few overlaps in the Classes, and the above numerical test helps finding them. The first one is between Classes 2 and 3 when $t=p-1$. These in fact look the same, but according to our bordering definitions, this case  ought to belong to Class 2. Similarly, there is an overlap between Classes 5 and 6 when $p=t=2$. Again they appear to be the same, but bordering definitions dictate that this configuration belongs to Class 6. In this way, the above numerical test allows us to identify the unique Class that each configuration belongs to.
\section{Discussion}

The authors in \cite{SB10} asked the question of what combinations of $d$, $t$, and $p$ lead to valid configurations of symmetrons. In this paper we give an answer to their question, and classify the the configurations of admissible icosahedral viruses. In particular,  Table \ref{coords} lists extra equations which must be fulfilled, and it is necessary that $d>0$ in non-Class 1 configurations, $t\ge0$ and $p>0$ in all configurations. For example, we can see that any combination of $t$ and $p$ will be a valid Class 1 configuration, which is the same conclusion as the one given in \cite{SB10} but avoids the calculation of limits they had to perform.

More interestingly, we can ask which solutions could be found when given $h$ and $k$ parameters. They could be multiple solutions for $d$, $t$, and $p$ and are subject to the aforementioned restrictions as well as the parity conditions on $h$ and $k$ (or the special conditions on $h$ and $k$ for Classes 5 and 6). Thus if $(h,k)\equiv(1,1)\mod2$, we may have solutions from Classes 1, 2, and 4. If $(h,k)\equiv(0,0)\mod2$, we may have solutions from Classes 2, 3, and 4 (and possibly 5 and 6). If $(h,k)\equiv(0,1)$ or $(1,0)\mod2$, we may have solutions from Classes 1, 3, and 4 (and possibly 6). Note that we can also have multiple solutions from the same Class. Furthermore, we can also use our knowledge of the restrictions and Table \ref{dtphkT} to find the exact conditions under which we will have solutions in any class.

One of the most simple and useful observations is that, in most cases, there will be up to 4 solutions, except in the cases of $(h,k)\equiv(0,0)\mod2$, when there will generally be up to 6 solutions, and of $h=k\equiv1\mod2$, when there will be 2 solutions. As an example, Table \ref{values} in Appendix \ref{Ap5} takes some values of $h$ and $k$ and lists all possible solutions of $d$, $t$, and $p$ and the Classes they belong to.

The formulas and parity conditions clearly show that the classes from \cite{SB10} are not simply degenerate cases of the ones we have found. Indeed, since we consider $d>0$, the disymmetron always creates an extra line of capsomers separating the two edges of pentasymmetrons of adjacent vertices. Moreover, the parity conditions are also consistent with previously known facts such as that even $T$ numbers are impossible without disymmetrons, as seen in  \cite{SB10} and \cite{W69}.
 
The solutions presented in this paper suggest the existence of certain configurations  of symmetrons, and as in the case of viruses with no disymmetrons \cite{SB10}, it would be interesting to consider whether the configurations presented here physically occur in reality. Moreover, since all   the cases are mathematically consistent, if those viruses may not exist in Nature, then further biological rules would need to be governing the behaviour of the viruses 
(in the discussion presented in \cite{SB10}, examples of features apparently favoured in Nature are given).

\begin{figure}[h]
\begin{center}
\includegraphics[width=0.45\textwidth]{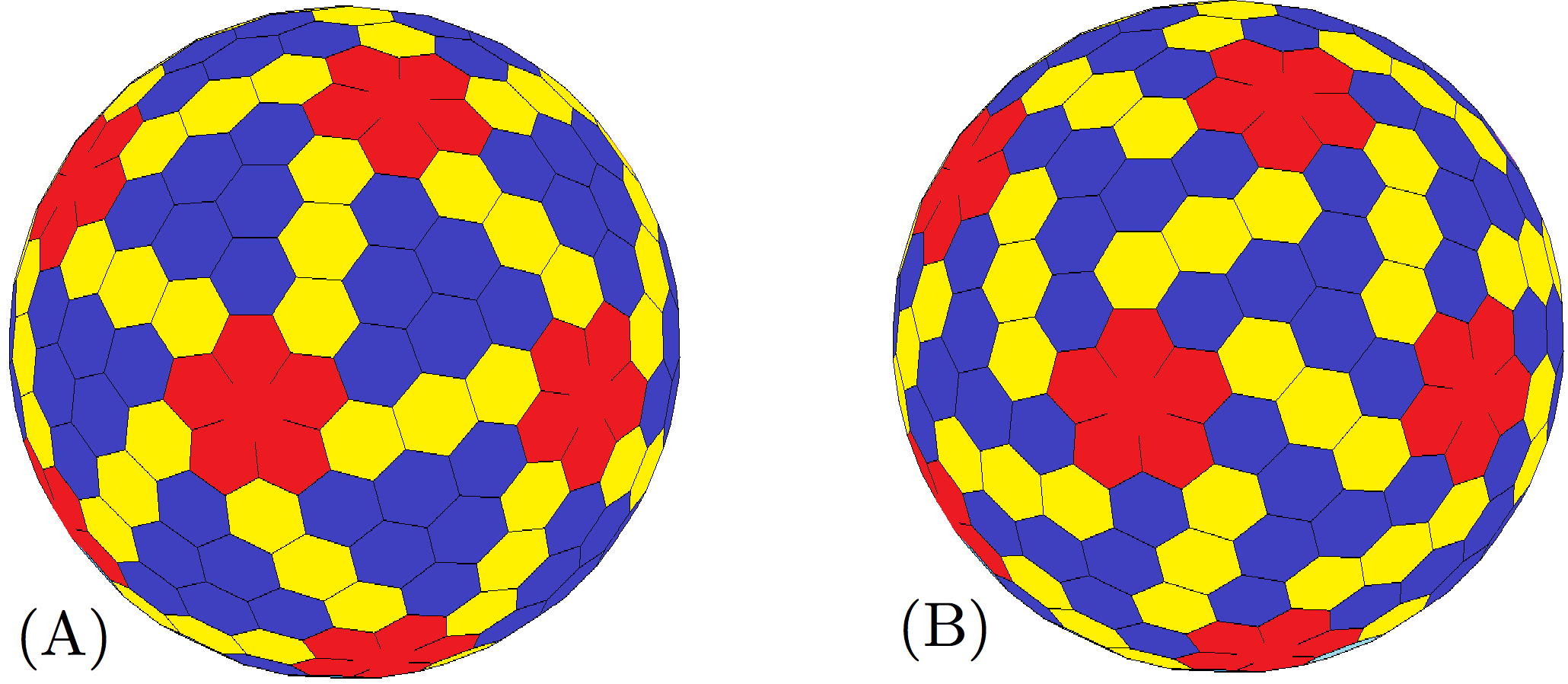}
\end{center}
\caption{The structure of the two known viruses with disymmetrons, which have $(h,k)=(2,4)$.  (A) Expected example: Class 3 virus with $d=3,~t=3,~p=2$. (B) The real, irregular virus with irregular trisymmetrons \cite{jaalinoja07}. } 
\label{fig:exceptionalex}
\end{figure}

Finally, whilst the viruses studied in this paper have regular structure and symmetrons, one may consider relaxing some of the mathematical constraints imposed, in order to classify non-regular viruses. This would be necessary in order to understand, for instance, the appearance of two known viruses P23-77 and SH1 that have disymmetrons \cite{happonen12,jaalinoja07}, and which have the same general structure, as depicted in Figure \ref{fig:exceptionalex}.A. 

The viruses P23-77 and SH1 have structure invariants  $(h,k)=(2,4)$, so the expected solutions would be Class 2 with edge lengths $(d,t, p)$ given by $(1,2,3)$; Class 2 with $(3,0,3)$; Class 3 with $(3,3,2)$ (shown in Figure \ref{fig:exceptionalex}.B); Class 4 with $(5,2,2)$; or Class 4 with $(5,3,1)$. These viruses do not fit into any of our solutions because of their irregular trisymmetrons, which are not equilateral triangles.  
 
\noindent {\bf Acknowledgments:} The research in this paper was conducted by the first author under the supervision of the latter, as part of the MIT PRIMES-USA program. Both authors would like to thank Robert Sinkovits for his enlightening comments, Steven Bradlow for inspiring conversations on the geometry of viruses  and MIT PRIMES-USA for the opportunity to conduct this research together.  The work of LPS is partially supported by NSF DMS-1509693.


 \appendix
 
 
 
\section{Proof that bordering symmetrons must exist if disymmetrons exist.}\label{Ap3}
 
 \begin{prop}
 Bordering symmetrons must exist if disymmetrons exist.
 \end{prop}
\textit{\textbf{Proof:}} We set the coordinate grid so that the disymmetron occupies the lattice points on $k=0$. Consider the adjacent $k=1$ lattice points. We will prove the statement by contradiction, so assume that no symmetrons border the disymmetron. Recalling which symmetrons may be adjacent to a disymmetron, the lack of bordering means that there can be at most 1 adjacent $k=1$ pentasymmetron capsomer, at most 1 adjacent $k=1$ trisymmetron capsomer, and at most 2 adjacent $k=1$ disymmetron capsomers, for a total of at most 4 adjacent $k=1$ capsomers, implying that $1\le d\le 3$.

If $d=1$: if disymmetrons are adjacent, then the 6-fold symmetry relating adjacent disymmetrons forces that $p=1$ and $t=0$, in which case bordering occurs. If disymmetrons are not adjacent, then the 6 points around the disymmetron must be occupied by capsomers of at most 2 pentasymmetrons and 2 trisymmetrons. By the Pigeonhole Principle, some adjacent symmetron has at least 2 adjacent capsomers, in which case bordering occurs.

If $d=2$, there are 3 adjacent capsomers on each side, and we can do casework based on what type of capsomer is in the middle. The middle capsomer can not be a pentasymmetron capsomer, or else the pentasymmetron would be bordering.
If the middle bordering capsomer is a disymmetron capsomer: without loss of generality, one may set one disymmetron at $(0,0)$ and $(1,0)$ and the other at $(0,1)$ and $(0,2)$. Consider the capsomer at $(1,1)$. This is a middle bordering capsomer to the disymmetron at $(0,1)$ and $(0,2)$, so it must be a disymmetron capsomer. This fixes $t=0$ and $p=1$, in which case bordering occurs.
Finally, if the middle bordering capsomer is a trisymmetron capsomer, the disymmetron is at $(0,0)$ and $(1,0)$. A trisymmetron has a vertex at $(0,1)$ and has a fixed orientation to avoid bordering. It has other vertices at $(0,t)$ and $(1-t,t)$. Moreover, 3-fold symmetry means a disymmetron exists from $(0,t+1)$ to $(1,t)$, and 6-fold symmetry relating adjacent disymmetrons fixes a pentasymmetron center at $(1+t,0)$. But this means that capsomers with $h=1$ and $1\le k\le t-1$ are unoccupied, and they can not be penta-, tri-, or disymmetrons. This contradiction implies that $t-1<1$, meaning $t=1$, which leads to bordering.

If $d=3$: since we assume no bordering occurs, among the adjacent $k=1$ capsomers, there must be 1 adjacent pentasymmetron capsomer (with $p>1$), 1 adjacent trisymmetron capsomer (with $t>1$), and 2 adjacent disymmetron capsomers. We set the disymmetron from $(1,0)$ to $(3,0)$. For $p>1$ to be true, the adjacent pentasymmetron capsomer must be at $(0,1)$ or $(3,1)$. Without loss of generality, we set it at $(0,1)$, so 2-fold symmetry means that another adjacent pentasymmetron capsomer is at $(4,-1)$.
If there were an adjacent disymmetron capsomer at $(1,1)$, then it would need to lie from $(1,1)$ to $(1,3)$ to be properly positioned with respect to 6-fold symmetry about the pentasymmetron containing $(0,1)$. However, by the 6-fold symmetry relating adjacent disymmetrons, this would force $p=1$, which is a contradiction and leads to bordering.
Therefore, the adjacent disymmetron capsomers must be at $(2,1)$ and $(3,1)$, and the trisymmetron bordering capsomer is at $(1,1)$. Because $t>1$, the trisymmetron also includes $(0,2)$ and $(1,2)$. This fixes the orientation of the disymmetron through $(2,1)$, which must go from $(2,1)$ to $(2,3)$. But this disymmetron will border the trisymmetron, which leads to a contradiction.
Thus, bordering must occur if $d>0$.\hfill $\square$

\section{On rotations in the $h,k$ grid.} \label{rotate60}\label{rotate60center}\label{rotate120center}\label{Ap1}
 
\begin{prop}A point $(x,y)$ rotated $\frac{\pi}{3}$ counterclockwise about the origin is mapped to $(-y,x+y)$.
\end{prop}
\textit{\textbf{Proof:}} A point $(x,y)$ can be interpreted as the vector $x\hat{h}+y\hat{k}$. If we denote the rotation with the function $R$ acting on some vector $\vec{v}$, then notice that $R$ is a linear function, i.e. $R(\vec{v_1}+\vec{v_2}+\dots+\vec{v_n})=R(\vec{v_1})+R(\vec{v_2})+\dots+R(\vec{v_n})$ and $R(a\vec{v})=aR(\vec{v})$. This means that 
$$R(x\hat{h}+y\hat{k})=xR(\hat{h})+yR(\hat{k}).$$ We can see that $R(\hat{h})=\hat{k}$ and $R(\hat{k})=\hat{k}-\hat{h}$. Therefore, $R(x\hat{h}+y\hat{k})=x\hat{k}+y\hat{k}-y\hat{h}$, which is  $(-y,x+y)$. \hfill $\square$\\

In particular, the above implies the following:

\begin{prop}A point $(x,y)$ rotated $120^\circ$ counterclockwise about the origin is mapped to $(-x-y,x)$.\end{prop}

By translating the grid so that $(x_0,y_0)$ becomes the origin, one takes $(x,y)\to(x-x_0,y-y_0)$, leading to the following:

 \begin{prop}A point $(x,y)$ rotated $\frac{\pi}{3}$ counterclockwise about $(x_0,y_0)$ is mapped to $(-y+x_0+y_0,x+y-x_0)$.\end{prop}

\textit{\textbf{Proof:}} Applying the above results, translation by $\frac{\pi}{3}$ counterclockwise takes $$(x-x_0,y-y_0)\to(y_0-y,x+y-x_0-y_0).$$ The statement follows by translating the grid back, $(y_0-y,x+y-x_0-y_0)\to(-y+x_0+y_0,x+y-x_0)$.\hfill$\square$\\

Similarly,  since 
\begin{eqnarray} 
  (x,y)&\rightarrow&(x-x_0,y-y_0)\nonumber\\
  &\rightarrow&(x_0+y_0-x-y,x-x_0)\nonumber\\
  &\rightarrow&(-x-y+2x_0+y_0,x-x_0+y_0).\nonumber
  \end{eqnarray}
 one has the following:
\begin{prop} A point $(x,y)$ rotated $120^\circ$ counterclockwise about the point $(x_0,y_0)$ is mapped to the point  $(-x-y+2x_0+y_0,x-x_0+y_0)$.  \end{prop} 

 \section{Adjacency of capsomers.}\label{diadj}\label{pentatriadj}\label{Ap2}
 
 \begin{prop}
 In the presence of disymmetrons, two capsomers from different pentasymmetrons or two capsomers from different trisymmetrons cannot be adjacent.\end{prop}
\textbf{\textit{Proof:}} In the following proof, the word ``symmetron" is used as shorthand for pentasymmetrons and trisymmetrons. By a symmetry of overlapping argument, if the two symmetrons have adjacent capsomers, then they are centered on adjacent vertices / faces of the icosahedron. Therefore, one symmetron can be mapped to the other by 2-fold symmetry about a 2-fold center. Because equilateral triangles and hexagons are convex and cannot contain the 2-fold center, two rays can be drawn from the 2-fold center such that the region between them, with an angle $<180^\circ$, contains a symmetron. If we reverse the direction of these two rays, they must bound the other symmetron by 2-fold symmetry. Therefore, we can draw a line $l$ through the 2-fold center that passes through neither of the bounded regions, so it separates the symmetrons. Because of the shape of these two symmetrons, we can always draw $l$ parallel to one of the lines $h=0$, $k=0$, or $h+k=0$.

The 2-fold center can either be a lattice point or halfway between two adjacent lattice points. In the first case, line $l$, which is along the triangular grid, separates the symmetrons and does not contain any capsomers of the symmetrons, making adjacency impossible. In the other case, the 2-fold center is between two adjacent lattice points. We orient the coordinate system so that the 2-fold center is $(\frac{1}{2},\frac{1}{2})$. As we saw in the first case, line $l$ can not be a line along the triangular grid for adjacency to occur, thus without loss of generality, we may assume that line $l$ has equation $k=\frac{1}{2}$. 

The disymmetron, which must contain the 2-fold center, must have capsomers at least at $(1,0)$ and $(0,1)$. Consider the symmetron with $k\ge1$. For adjacency to occur, this symmetron must have capsomers with $k=1$ and $h<0$ or $h>0$, and the opposite symmetron must have capsomers with $k=0$ and $h<0$ or $h>0$, respectively. Moreover, 2-fold symmetry dictates that the $k\ge1$ symmetron has capsomers with $k=1$ and $h>0$ or $h<0$, respectively. So this symmetron has $k=1$ capsomers on both sides of $h=0$. But the disymmetron lies at $(0,1)$. Thus, the convexity of the symmetrons makes this impossible and the statement follows.\hfill $\square$

\section{On the triangulation number $T$}\label{Ap4}

Seen as vectors, the distance squared from $(0,0)$ to $(x,y)$ is $x^2+xy+y^2$,
%
%
which explains why $T=h^2+hk+k^2$:   turning the $k$-axis so that it is perpendicular to the $h$-axis, as with rectangular coordinates,  $h^2+hk+k^2$ gives the number of unites in the square with side length equal to an icosahedral edge. However, if we want to find the area of an icosahedral face, we only want half of this square.  Since we want to count triangles, 
$h^2+hk+k^2$ also gives the number of triangles per icosahedral face.
 \newpage
\section{Table of solutions for some   $h$ and $k$}\label{Ap5}

\begin{table}[h]

\scalebox{0.8}{
\begin{tabular}{|c|c|}
 \hline
 $(h,k)$ & Class:$(d,t,p)$\\\hline
(0,1) & 1:(0,0,1)\\\hline
(0,2) & 2:(1,0,1)\\\hline
(0,3) & 1:(0,1,2), 3:(2,1,1)\\\hline
(0,4) & 3:(3,2,1), 4:(3,0,2), 6:(1,2,2)\\\hline
(0,5) & 1:(0,2,3), 3:(4,3,1), 6:(2,3,2)\\\hline
(0,6) & 3:(5,4,1), 4:(5,1,3), 6:(3,4,2)\\\hline
(0,7) & 1:(0,3,4), 3:(6,5,1), 6:(4,5,2)\\\hline
(0,8) & 3:(7,6,1), 4:(7,2,4), 6:(5,6,2)\\\hline
(1,1) & 1:(0,1,1)\\\hline
(1,2) & 1:(0,2,1), 1:(0,0,2), 4:(2,0,1)\\\hline
(1,3) & 1:(0,3,1), 1:(0,2,2), 2:(2,0,2)\\\hline
(1,4) & 1:(0,4,1), 1:(0,1,3), 4:(4,1,2)\\\hline
(1,5) & 1:(0,5,1), 1:(0,3,3), 4:(4,0,3), 5:(2,2,3)\\\hline
(1,6) & 1:(0,6,1), 1:(0,2,4), 4:(6,2,3)\\\hline
(1,7) & 1:(0,7,1), 1:(0,4,4), 4:(6,1,4)\\\hline
(1,8) & 1:(0,8,1), 1:(0,3,5), 4:(8,3,4)\\\hline
(2,2) & 2:(1,1,2), 4:(3,1,1)\\\hline
(2,3) & 1:(0,3,2), 1:(0,0,3), 3:(2,2,2), 4:(4,2,1)\\\hline
(2,4) & 2:(1,2,3), 2:(3,0,3), 3:(3,3,2), 4:(5,2,2), 4:(5,3,1)\\\hline
(2,5) & 1:(0,4,3), 1:(0,1,4), 3:(4,4,2), 4:(6,4,1)\\\hline
(2,6) & 2:(1,3,4), 3:(5,5,2), 4:(7,3,3), 4:(5,0,4), 4:(7,5,1), 5:(3,2,4)\\\hline
(2,7) & 1:(0,5,4), 1:(0,2,5), 3:(6,6,2), 4:(8,6,1)\\\hline
(2,8) & 2:(1,4,5), 3:(7,7,2), 4:(9,4,4), 4:(7,1,5), 4:(9,7,1)\\\hline
(3,3) & 1:(0,4,2), 2:(2,1,3)\\\hline
(3,4) & 1:(0,5,2), 1:(0,0,4), 3:(2,3,3), 4:(6,3,2)\\\hline
(3,5) & 1:(0,6,2), 1:(0,5,3), 2:(2,2,4), 2:(4,0,4)\\\hline
(3,6) & 1:(0,7,2), 1:(0,1,5), 3:(2,4,4), 4:(8,4,3)\\\hline
(3,7) & 1:(0,8,2), 1:(0,6,4), 2:(2,3,5), 4:(6,0,5), 5:(4,2,5)\\\hline
(3,8) & 1:(0,9,2), 1:(0,2,6), 3:(2,5,5), 4:(10,5,4)\\\hline
(4,4) & 2:(3,1,4), 3:(3,4,3), 4:(7,4,2)\\\hline
(4,5) & 1:(0,6,3), 1:(0,0,5), 3:(4,5,3), 4:(8,5,2)\\\hline
(4,6) & 2:(3,2,5), 2:(5,0,5), 3:(3,5,4), 3:(5,6,3), 4:(9,5,3), 4:(9,6,2)\\\hline
(4,7) & 1:(0,7,4), 1:(0,1,6), 3:(6,7,3), 4:(10,7,2)\\\hline
(4,8) & 2:(3,3,6), 3:(3,6,5), 3:(7,8,3), 4:(11,6,4), 4:(7,0,6), 4:(11,8,2), 5:(5,2,6)\\\hline
(5,5) & 1:(0,7,3), 2:(4,1,5)\\\hline
(5,6) & 1:(0,8,3), 1:(0,0,6), 3:(4,6,4), 4:(10,6,3)\\\hline
(5,7) & 1:(0,9,3), 1:(0,8,4), 2:(4,2,6), 2:(6,0,6)\\\hline
(5,8) & 1:(0,10,3), 1:(0,1,7), 3:(4,7,5), 4:(12,7,4)\\\hline
(6,6) & 2:(5,1,6), 3:(5,7,4), 4:(11,7,3)\\\hline
(6,7) & 1:(0,9,4), 1:(0,0,7), 3:(6,8,4), 4:(12,8,3)\\\hline
(6,8) & 2:(5,2,7), 2:(7,0,7), 3:(5,8,5), 3:(7,9,4), 4:(13,8,4), 4:(13,9,3)\\\hline
(7,7) & 1:(0,10,4), 2:(6,1,7)\\\hline
(7,8) & 1:(0,11,4), 1:(0,0,8), 3:(6,9,5), 4:(14,9,4)\\\hline
(8,8) & 2:(7,1,8), 3:(7,10,5), 4:(15,10,4)\\\hline
\end{tabular}
}

\caption{For some values of $h$ and $k$, all possible solutions of $d$, $t$, and $p$ and the Classes they belong to.}
   \label{values}
 \end{table}

\begin{small}
  
 \end{small}

\end{document}